\documentclass[twocolumn,epsf,showpacs,prl]{revtex4}
\usepackage{graphicx}
\usepackage{dcolumn}
\usepackage{bm}

\begin{document}

\title{Gamma Ray Astronomy with Magnetized Zevatrons}
\author{Eric Armengaud$^{a,b}$, G{\"u}nter Sigl$^{a,b}$, Francesco Miniati$^c$}

\affiliation{$^a$APC~\footnote{UMR 7164 (CNRS, Universit\'e Paris 7, CEA, Observatoire de Paris)} (AstroParticules et Cosmologie),
11, place Marcelin Berthelot, F-75005 Paris, France}

\affiliation{$^b$ GReCO, Institut d'Astrophysique de Paris, C.N.R.S.,
98 bis boulevard Arago, F-75014 Paris, France}

\affiliation{$^c$ Physics Department, ETH Z\"urich, 8093 Z\"urich, 
Switzerland}

\begin{abstract}
  Nearby sources of cosmic rays up to a ZeV($=10^{21}\,$eV) could be
  observed with a multi-messenger approach including secondary
  $\gamma-$rays and neutrinos. If cosmic rays above $\sim10^{18}\,$eV
  are produced in magnetized environments such as galaxy clusters, the
  flux of secondary $\gamma-$rays below $\sim1\,$TeV can be enhanced
  up to several orders of magnitudes compared to unmagnetized sources.
  A particular source of enhancement are synchrotron and cascade
  photons from $e^+e^-$ pairs produced by protons from sources with
  relatively steep injection spectra $\propto E^{-2.6}$. Such sources
  should be visible at the same time in ultra-high energy cosmic ray
  experiments and $\gamma-$ray telescopes.
\end{abstract}


\maketitle

{\it Introduction.}
One of the central unresolved issues of modern astroparticle physics
is the origin of cosmic rays, particularly those at Ultra High Energy
(UHECR) which have been observed at energies up to a few times
$10^{20}\,$eV~\cite{review}. Sources capable of accelerating such particles,
 like powerful radio galaxies commonly found inside galaxy clusters and groups, 
 are thought to be rare~\cite{torres-review} and have yet to be identified.

Astroparticle physics is currently experimentally driven and involves
many different existing or planned projects ranging from UHECR
observatories such as the Pierre Auger project~\cite{auger}, to neutrino
telescopes~\cite{nu_review}, as well as ground and space based $\gamma-$ray
detectors operating at TeV and GeV energies,
respectively~\cite{gammarev}. It is clear that GeV-TeV $\gamma-$ray
and neutrino astronomy will prove an invaluable tool to unveil the
sources, and probe into the mechanism, of UHECRs.  Even if a putative
source were to produce exclusively UHECRs, photo-pion~\cite{gzk} and
pair production by protons on the cosmic microwave background (CMB)
would lead to a guaranteed secondary photon and neutrino fluxes that
could be detectable.

Secondary photon fluxes from UHECR interactions with the CMB have been
discussed before in the
literature~\cite{Gabici:2005gd,Ferrigno:2004am,Rordorf:2004jp}.
In these works, however, photo-pair production by protons, photons
generated by the GZK interaction and the effect of structured magnetic
fields on both the electromagnetic (EM) cascades and UHECRs were not
taken into account in a quantitative and comprehensive way.
In Ref.~\cite{Inoue:2005vz} proton acceleration up to
$\sim10^{19}\,$eV around cluster accretion shocks was studied which,
however, can not explain the highest energy cosmic rays. In the present
{\it letter} we find that in combination, these ingredients can
increase secondary photon fluxes from individual UHECR sources below
$\sim1\,$TeV up to several orders of magnitude. This potentially makes
their detection much easier.

Furthermore, we extend the investigation to also consider the case of
steep proton spectra $\propto E^{-2.6}$ at ultra-high energies. This
is motivated by the scenario where extragalactic protons dominate the
observed flux down to the ``second knee'' at $\simeq4\times
10^{17}\,$eV, such that the ankle at $\simeq5\times10^{18}\,$eV is
caused by photopair production by the extragalactic
protons~\cite{Berezinsky:2002nc,Berezinsky:2005cq}. 

In the following we compute the expected $\gamma$-ray flux from $\sim 10$ MeV
to the highest energies, by combining simulations of UHECR propagation
in structured large scale extragalactic magnetic
fields~\cite{Sigl:2004yk} (EGMF) with production of secondary hadrons
by nucleon primaries using the event generator SOPHIA~\cite{sophia}.
Pair production by protons on the CMB is taken into account as a
continuous energy loss: a proton with energy $E$ generates 
electron-positron pairs (heretoafter simply referred to as electrons) 
with a power-law energy distribution $dn/dE_e\propto E_e^{-7/4}$
for $E_e\lesssim E$~\cite{mastichiadis}. All the electromagnetic
products of these interactions are then followed to the observer using
an EM cascade code based on Ref.~\cite{Lee:1996fp}, which takes into
account the inhomogeneous distribution of magnetic fields in
the simulation box and the presence of a cosmic infrared background
from~\cite{Primack:1998wn}.

We consider the case of a discrete source in one of the prominent
magnetized galaxy clusters from the simulations based on
Refs.~\cite{ryu,miniati}, with a size comparable to the VIRGO
cluster. Sources in such clusters could be active galaxies (AGNs)
whose hot spots, for example, have been suggested to accelerate
UHECRs~\cite{torres-review}. We use $c=1$ throughout the letter.

{\it Influence of Pair Production and Proton Injection Spectra on GeV-TeV
Photon Fluxes.}
We assume point sources whose flux contributes a fraction $\eta\leq1$
to the total solid integrated UHECR flux observed around $10^{19}\,$eV,
i.e. $\simeq2.5\times10^{-19}(\eta/0.01)\,{\rm cm}^{-2}\,{\rm
s}^{-1}$, the approximate average over existing flux measurements~\cite{review}.
For distances $d\lesssim500\,$Mpc this flux is not too much influenced by
energy losses and roughly correponds to an UHECR injection power above
$10^{19}\,$eV of
\begin{equation}
  L_{19}\simeq4.8\times10^{42}\left(\frac{d}{100\,{\rm Mpc}}\right)^2
  \left(\frac{\eta}{0.01}\right)\,{\rm erg\,s}^{-1}\,.\label{eps}
\end{equation}

\begin{figure}[ht]
\begin{center}
\includegraphics[width=0.48\textwidth]{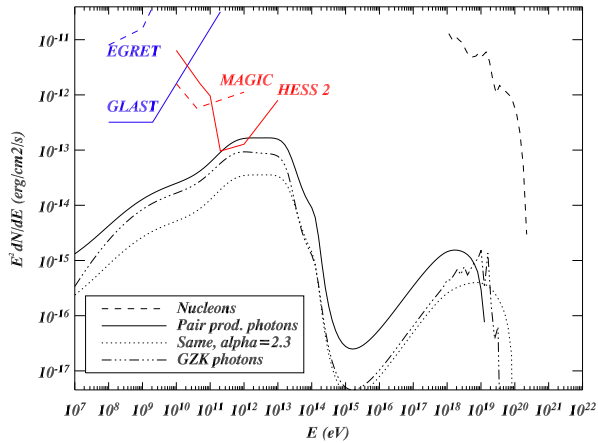}
\caption{Comparison of differential $\gamma$-ray fluxes (multiplied by squared
energy) generated by GZK interactions and photo-pair production for a source at
100 Mpc distance injecting protons up to 1 ZeV with a spectrum $\propto E^{-2.7}$
(solid and dashed-dotted lines) and with a spectrum $\propto E^{-2.3}$ (dotted
line) in the absence of magnetic fields. The power emitted above
$10^{19}\,$eV is $L_{19} = 10^{43}\,$erg s$^{-1}$, corresponding to
$\eta\simeq0.02$ in Eq.~(\ref{eps}). Also shown are point flux sensitivities
of the $\gamma-$ray instruments EGRET~\cite{egret}, GLAST~\cite{glast}, HESS 2~\cite{hess}, and MAGIC~\cite{magic}. The dashed line on
the right is the observable primary UHECR flux from the source.}\label{fig1}
\end{center}
\end{figure}

Fig.~\ref{fig1} demonstrates the contribution of pair production to the
photon fluxes at GeV and TeV energies. It increases with the steepness of the
proton injection spectrum because pair production is the dominant
energy loss process for protons with energies
$10^{18}\lesssim E \lesssim 4\times 10^{19}\,$eV. It appears that, for steep enough
injection spectrum $\propto E^{-\alpha}$ with $\alpha \sim 2.6 -2.7$, necessary
if extragalactic cosmic rays dominate above the
ankle~\cite{Berezinsky:2002nc,Berezinsky:2005cq}, and large enough
distances between the source and the observer, photo-pair production
dominates over pion production for the secondary $\gamma$-ray flux.

We now investigate the consequences of the low energy extension of the
UHECR injection spectrum.

Cosmic ray protons of energy $E$ and integral flux $J^{\rm l}_{\rm
  CR}(E)$ confined within a volume $\simeq R^3$ which interact with a
baryon gas of density $n_b$ produce $\gamma-$rays of energy $\simeq
f_\gamma E\simeq0.1E$ at a rate $\simeq J^{\rm l}_{\rm
  CR}(E)\sigma_{pp}\,n_b\,R^3$. Here, the proton-proton cross section
$\sigma_{pp}\simeq3\times10^{-26}\,{\rm cm}^2$ can be approximated as
energy independent. For proton confinement times $t_{\rm
  conf}(E)\gtrsim R$, we can express the confined integral flux
in terms of the total injection rate $I^{\rm inj}_{\rm CR}(E)$ as
$J^{\rm l}_{\rm CR}(E)\simeq I^{\rm inj}_{\rm CR}(E)\,t_{\rm
  conf}(E)/R^3$. Furthermore, in a steady state situation, the leaking
cosmic ray flux observed at a distance $d$ is $J^{\rm obs}_{\rm
  CR}(E)\simeq M(E)\,I^{\rm inj}_{\rm CR}(E)/(4\pi d^2)$, where
$M(E)\lesssim1$ is a modification factor accounting for interaction
losses during propagation to the observer. We can then relate the
integral photon flux at energy above $E_\gamma$ from $pp$ interactions
within the volume $R^3$, $J^{pp}_\gamma(E_\gamma)$, to the integral
UHECR flux above $E_{\rm CR}$, $J^{\rm obs}_{\rm CR}(E_{\rm CR})$,
\begin{equation}
  J^{pp}_\gamma(E_\gamma)\simeq\sigma_{pp}n_b\,t_{\rm conf}(E_\gamma/f_\gamma)
  \frac{I^{\rm inj}_{\rm CR}(E_\gamma/f_\gamma)}{I^{\rm inj}_{\rm CR}(E_{\rm CR})}
  \,\frac{J^{\rm obs}_{\rm CR}(E_{\rm CR})}{M(E_{\rm CR})}\,.\label{J_egret}
\end{equation}
Both $J^{pp}_\gamma$ and $J^{\rm obs}_{\rm CR}(E_{\rm CR})$ are fluxes observed at distance $d$.

For a galaxy cluster $n_b\sim10^{-3}\,{\rm cm}^{-3}$,
$R\simeq2\,$Mpc and $t_{\rm conf}(E_\gamma/f_\gamma)\lesssim10^{10}\,$yr
(the age of the Universe). Thus, the optical depth for $pp$
interaction is $\sigma_{pp}n_b\,t_{\rm conf}\lesssim0.3$. Furthermore,
at $E_{\rm CR}\simeq10^{19}\,$eV, $M(E_{\rm CR})\simeq1$ and with the cosmic
ray flux from Eq.~(\ref{eps}) we have
\begin{equation}
  J^{pp}_\gamma(E_\gamma)\simeq7.5\times10^{-20}\,\left(\frac{\eta}{0.01}\right)\,
  \frac{I^{\rm inj}_{\rm CR}(E_\gamma/f_\gamma)}
  {I^{\rm inj}_{\rm CR}(10^{19}\,{\rm eV})}\,
  {\rm cm}^{-2}\,{\rm s}^{-1}\,.\label{J_egret2}
\end{equation}

The upper limit on the $\gamma$-ray flux at $E_\gamma\sim100\,$MeV
from EGRET is typically $\sim 4 \times 10^{-8}\,{\rm cm}^{-2}\,{\rm
s}^{-1}$ for clusters like Coma or Virgo~\cite{egret}. 
Eq.~(\ref{J_egret2}) thus implies the condition
\begin{equation}
  \frac{I^{\rm inj}_{\rm CR}(E_\gamma/f_\gamma)}
  {I^{\rm inj}_{\rm CR}(10^{19}\,{\rm eV})}\lesssim 5.3\times 10^{11}
  \left(\frac{0.01}{\eta}\right)\,.\label{egret_constr}
\end{equation}
For an unbroken power law $I^{\rm inj}_{\rm CR}(E)\propto E^{1-\beta}$
for $E_\gamma/f_\gamma\lesssim E\lesssim 10^{19}\,$eV, this would imply
the relatively strong constraint

\begin{equation}
\beta\lesssim2.17-0.1\log(\eta/0.01).
\label{beta_constraint}
\end{equation}

\noindent This constraint can be avoided
if the power law cuts off or becomes harder at low energies such that
Eq.~(\ref{egret_constr}) is satisfied.  In particular, in the scenario
in which extragalactic protons dominate down to a few $10^{17}\,$eV,
their injection spectrum,
$\alpha\simeq2.6$~\cite{Berezinsky:2002nc,Berezinsky:2005cq}, cannot
continue below $\sim10^{11+\log(\eta/0.01)/1.6}\,$eV.

At the same time, for $\beta\geq2.$, the total power emitted by the
source in cosmic rays down to energy $E_{\rm CR}^{\rm min}$ is $L_{\rm
  CR}\simeq \left(10^{19}\,{\rm eV}/ E_{\rm CR}^{\rm
    min}\right)^{\beta-2}\,L_{19}$. Therefore, low energy cosmic ray flux
extensions with power law index not much larger than 2
also assure reasonable total cosmic ray powers which remain largely below the
high end of bolometric luminosity for AGNs, $L\lesssim10^{48}\,{\rm
  erg}\,{\rm s}^{-1}$.

\begin{figure}[ht]
\begin{center}
\includegraphics[width=0.48\textwidth]{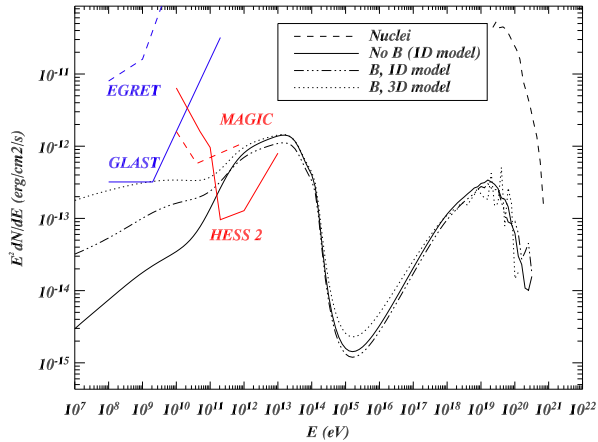}
\caption{Differential $\gamma$-ray fluxes (multiplied by squared energy) from photo-pion and pair production by UHECR injected with an $E^{-2.7}$ spectrum by a source at 20 Mpc. We assume $\eta \simeq 0.3$, corresponding to a proton luminosity $L_{19} \simeq 7 \times 10^{42}\,$erg s$^{-1}$ in Eq.~(\ref{eps}). Compared are different magnetic field and propagation models, as indicated.}
\label{fig2}
\end{center}
\end{figure}

{\it Sources in Magnetized Galaxy Clusters.}
Fig.~\ref{fig2} demonstrates the influence of the cluster EGMF on the
fluxes of secondary $\gamma-$rays. Note that
all fluxes scale with $\eta$. This implies that
for magnetized galaxy clusters and relatively soft injection spectrum
$\propto E^{-2.7}$, the TeV $\gamma-$ray signal should be visible
at least with HESS 2, provided $\eta\gtrsim0.02$, whereas detectability
by MAGIC and GLAST requires $\eta\gtrsim0.3$. For harder injection
spectra $\propto E^{-2.3}$ these numbers are $\eta\gtrsim0.05$ and
$\eta\gtrsim0.5$, respectively. These figures are for a source at $d=20\,$Mpc, but
depend only moderately on $d$.

Currently the isotropy of UHECRs at $10^{19}$ eV imposes only loose
bounds on $\eta$ due to the small statistics. Upcoming generation
experiments like Auger will constrain $\eta$ much better in a near future.

Fig.~\ref{fig2} shows that the total amount of energy going into $\gamma-$rays
is roughly 10\% of the energy in protons  above the GZK threshold and
does not depend much on the intracluster magnetic field. 
This is because, except for very steep injection spectra, most of the energy
going into photons is due to pion production whose mean free path
above $\simeq6\times10^{19}\,$eV is already smaller than the source
distance. In contrast, the tail below $\simeq1\,$TeV is due to
interactions of cosmic-rays below the GZK threshold whose energy loss
length is much larger than the source distance such that their path length
can be considerably increased by the intracluster magnetic field.
At these energies the dominant emission mechanism is 
synchrotron radiation from $e^+e^-$ pairs produced by UHECR protons,
as we will see in the following.

Electrons of energy $E$ in a magnetic field 
$B$ emit synchrotron photons of typical energy
\begin{equation}
  E_{\rm syn}\simeq6.8\times10^{11}
  \left(\frac{E}{10^{19}\,{\rm eV}}\right)^2
  \left(\frac{B}{0.1\,\mu{\rm G}}\right)\,{\rm eV}
  \,.\label{esynch}
\end{equation}
The typical energy of electrons and photons produced in pion
production is $\sim5\times10^{18}\,$eV~\cite{gzk}, whereas most of the
pair production occurs for proton energies $E$ between
$\simeq10^{18}\,$eV and $\simeq4\times10^{19}\,$eV, and gives rise to 
an electron energy distribution $dn/dE_e\propto E_e^{-7/4}$.
Therefore, in a $0.1\mu$G field, the synchrotron emission from the
electrons produced in the first stages of an EM cascade initiated by
pion production occurs mainly below $\sim0.1\,$TeV. For pair
production, if the proton injection spectra are steeper than $E^{-2}$,
most of the EM energy is produced by protons of a few times
$10^{18}\,$eV which end up in synchrotron photons below $\sim1\,$TeV
with a long tail to lower energies due to the rather flat pair
spectrum. Both these effects are seen in Fig.~\ref{fig2}.

The TeV photon flux in Figs.~\ref{fig1} and~\ref{fig2}
can be approximated as
\begin{equation}
  J^{\rm ph}_\gamma(E_\gamma\sim1\,{\rm TeV})\sim6.3\times10^{-14}
  \left(\frac{\eta}{0.03}\right)\,{\rm cm}^{-2}{\rm s}^{-1}\,.\label{J_pp}
\end{equation}
Requiring the TeV $\gamma-$ray flux to be dominated by the UHECR
interactions rather than low energy $pp$ interactions,
$J^{pp}_\gamma(\sim{\rm TeV})\lesssim J^{\rm ph}_\gamma(\sim{\rm TeV})$,
amounts to the condition $\beta \lesssim 1.95$.
As a consequence, in order for both the EGRET constraint and a TeV
$\gamma-$ray flux to be dominated by UHECR interactions rather than
low energy $pp$ interactions, would require a hard cosmic ray
injection spectrum below ultra-high energies.

In Fig.~\ref{fig2} the one-dimensional simulation neglected proton
deflection and used the 3d profile of the magnetic field projected
onto the line of sight. The 3d structure of the EGMF tends to
enhance the photon flux from pair production considerably.
This is because UHECR between the pair
production threshold at $\sim10^{18}\,$eV and the pion production
threshold can diffuse transverse to the line of sight producing photons
at $\sim$ GeV energies for timescales much longer than the rectilinear 
propagation time.  For a nearby source 
this would result in a substantial enhancement of this photon flux.
\begin{figure}[ht]
\begin{center}
\includegraphics[width=0.48\textwidth]{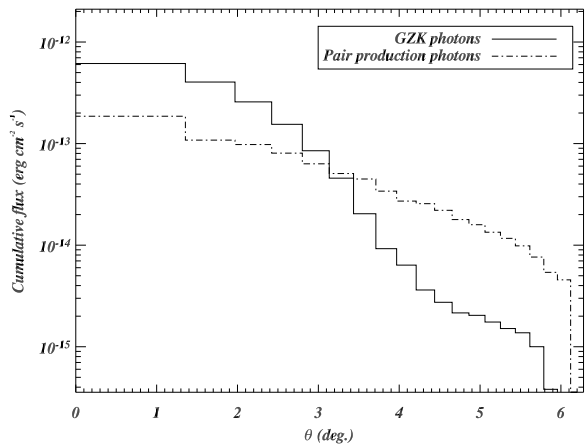}
\caption{Spatial extension of the counterpart in $\gamma-$ray above
a TeV of the magnetized source at 20 Mpc in the 3d model of Fig.~\ref{fig2}. The relative contributions of pair production and GZK photons are shown separately as cumulative fluxes emitted at off-sets from the source center larger than
$\theta$.}\label{fig3}
\end{center}
\end{figure}
This effect could also lead to a GeV-TeV $\gamma-$ray halo whose structure
could be observable in the case of a nearby powerful source 
with steep UHECR injection spectrum. In Fig.~\ref{fig3},
the cumulative flux $\int_{\theta}^{\infty}d\Omega E^2 dN/dE d\Omega$
is represented for a source in a magnetized cluster at 20 Mpc from the
observer. In this case the $\gamma-$ray halo has a spatial extension
of order $3^{\circ}$ and is dominated by pair production at angles
$\gtrsim3^\circ$.  Such a source located at 100 Mpc (like Coma) would
have an extension $\sim 0.6^{\circ}$, still resolvable by imaging
atmospheric \v{C}erenkov detectors.

GeV-PeV cosmic-ray protons and TeV electrons accelerated at cluster
shocks in galaxy clusters can also produce diffuse $\gamma-$ray
emission at a comparable level through $pp$ and inverse Compton emissions 
respectively~\cite{miniati03}.
The radiation spectrum produced by these mechanisms is a flat power
law, $E^2dN/dE\propto E^{-\alpha}$ with $\alpha\sim 0$. 
Thus it should be distinguishable from the spectra
illustrated in Fig.~\ref{fig2}, characterized by a broken power law
with $\alpha <0$ at TeV energies. Notice that the latter
is rather insensitive to the slope of the injected UHECRs as the 
emitting particles are produced in a cascade process.

Finally, the photon fluxes do not depend significantly on $E_{\rm max}$ in
our scenarios, provided $E_{\rm max}\gtrsim\,$few $10^{20}\,$eV.

{\it Conclusions.}
Ultra-high energy cosmic rays produce secondary $\gamma-$ray from pion
production and pair production on the cosmic microwave and other low energy
photon backgrounds. If a significant fraction of highest energy cosmic rays
is produced in galaxy clusters which are known to contain magnetic fields of fractions of a microGauss over Mpc length scales, the secondary $\gamma-$ray below
$\sim1\,$TeV could be detectable by $\gamma-$ray experiments such as HESS 2,
and potentially also by MAGIC and GLAST. This is especially the case for relatively steep injection spectra $\propto E^{-2.6}$ above $10^{18}\,$eV which are required by scenarios explaining the
ankle by pair production of extragalactic protons. Injection spectra
steeper than $\propto E^{-2.2}$, however, cannot continue down to $\sim$GeV
energies. Instead, they have to become harder somewhere between
GeV and ultra-high energies to avoid over-production of photons produced
in inelastic $pp$ collisions.

Whereas the $\gamma-$ray flux around a TeV turns out to be relatively insensitive to the magnetic field, its slope below $\sim\,$TeV contains information about
the cluster magnetic field and the ultra-high energy cosmic ray injection
spectrum. Furthermore the $\gamma-$ray flux is expected to extend over
the size of the magnetized region embedding the UHECR source, and the
TeV source could therefore be spatially resolved.

{\it Acknowledgements.} FM acknowledges support by the Swiss Institute of
Technology through a Zwicky Prize Fellowship.


\begin{thebibliography}{999}

\bibitem{review}
  J.~W.~Cronin,
  Nucl.\ Phys.\ Proc.\ Suppl.\  {\bf 138}, 465 (2005)

\bibitem{torres-review}
D.~F.~Torres and L.~A.~Anchordoqui,
Rept.\ Prog.\ Phys.\  {\bf 67}, 1663 (2004)

\bibitem{auger} J.~W.~Cronin, Nucl.~Phys.~B (Proc.~Suppl.) {\bf 28B} (1992)
213; The Pierre Auger Observatory Design Report (ed.~2), March 1997;
see also {\sf http://www.auger.org}.

\bibitem{nu_review} for recent reviews see, e.g.,
F.~Halzen and D.~Hooper,
Rept.\ Prog.\ Phys.\  {\bf 65}, 1025 (2002)
A.~B.~McDonald, C.~Spiering, S.~Schonert, E.~T.~Kearns and T.~Kajita,
Rev.\ Sci.\ Instrum.\  {\bf 75}, 293 (2004)

\bibitem{gammarev} for recent short reviews see, e.g.,
H.~J.~V\"olk,
arXiv:astro-ph/0401122;
H.~J.~V\"olk,
arXiv:astro-ph/0312585.

\bibitem{gzk}
K.~Greisen,
Phys.\ Rev.\ Lett.\  {\bf 16}, 748 (1966);
G.~T.~Zatsepin and V.~A.~Kuzmin,
JETP Lett.\  {\bf 4}, 78 (1966)
[Pisma Zh.\ Eksp.\ Teor.\ Fiz.\  {\bf 4}, 114 (1966)].

\bibitem{Gabici:2005gd}
  S.~Gabici and F.~A.~Aharonian,
  arXiv:astro-ph/0505462.

\bibitem{Ferrigno:2004am}
  C.~Ferrigno, P.~Blasi and D.~De Marco,
  Astropart.\ Phys.\  {\bf 23}, 211 (2005)

\bibitem{Rordorf:2004jp}
  C.~Rordorf, D.~Grasso and K.~Dolag,
  Astropart.\ Phys.\ {\bf 22}, 167 (2004)

\bibitem{Inoue:2005vz}
  S.~Inoue, F.~A.~Aharonian and N.~Sugiyama,
  Astrophys.\ J.\  {\bf 628}, L9 (2005)

\bibitem{Berezinsky:2002nc}
  V.~Berezinsky, A.~Z.~Gazizov and S.~I.~Grigorieva,
  arXiv:hep-ph/0204357.

\bibitem{Berezinsky:2005cq}
  V.~Berezinsky, A.~Z.~Gazizov and S.~I.~Grigorieva,
  ``Dip in UHECR spectrum as signature of proton interaction with CMB,''
  Phys.\ Lett.\ B {\bf 612}, 147 (2005)

\bibitem{Sigl:2004yk}
  G.~Sigl, F.~Miniati and T.~A.~Ensslin,
  Phys.\ Rev.\ D {\bf 70}, 043007 (2004)

\bibitem{sophia}
  A.~Mucke, R.~Engel, J.~P.~Rachen, R.~J.~Protheroe and T.~Stanev,
  Comput.\ Phys.\ Commun.\  {\bf 124}, 290 (2000)

\bibitem{mastichiadis} A.~Mastichiadis,
Mon.\ Not.\ Roy.\ Astron.\ Soc.\  {\bf 253}, 235 (1991).

\bibitem{Lee:1996fp}
  S.~Lee,
  Phys.\ Rev.\ D {\bf 58}, 043004 (1998)

\bibitem{Primack:1998wn}
J.~R.~Primack et al.,
Astropart.\ Phys.\ {\bf 11}, 93 (1999)

\bibitem{ryu} D.~Ryu, H.~Kang, and P.~L.~Biermann,
Astron.~Astrophys. {\bf 335} (1998) 19.

\bibitem{miniati}
F.~Miniati,
Mon.\ Not.\ Roy.\ Astron.\ Soc.\  {\bf 337}, 199 (2002)

\bibitem{egret}
O.~Reimer, M.~Pohl, P.~Sreekumar and J.~R.~Mattox,
Astrophys.\ J.\  {\bf 588}, 155 (2003)

\bibitem{glast} see, e.g., {\sf http://www-glast.stanford.edu}

\bibitem{hess} see {\sf http://www.mpi-hd.mpg.de/hfm/HESS/HESS.html}

\bibitem{magic} see, e.g., {\sf http://magic.mppmu.mpg.de/}

\bibitem{miniati03}
F.~Miniati,
Mon.\ Not.\ Roy.\ Astron.\ Soc.\  {\bf 342}, 1009 (2003)

\end{thebibliography}
\end{document}